# A Robust Missing Value Imputation Method MiFoImpute for Incomplete Molecular Descriptor Data And Comparative Analysis With Other Missing Value Imputation Methods

Doreswamy[1] and Chanabasayya .M. Vastrad[2]

[1]Department of Computer Science Mangalore University, Mangalagangotri-574 199, Karnataka, INDIA
doreswamyh@yahoo.com

[2]Department of Computer Science Mangalore University, Mangalagangotri-574 199, Karnataka, INDIA
channu.vastrad@gmail.com

## ABSTRACT

*Missing data imputation is an important research topic in data mining. Large-scale Molecular descriptor data may contains missing values (MVs). However, some methods for downstream analyses, including some prediction tools, require a complete descriptor data matrix. We propose and evaluate an iterative imputation method MiFoImpute based on a random forest. By averaging over many unpruned regression trees, random forest intrinsically constitutes a multiple imputation scheme. Using the NRMSE and NMAE estimates of random forest, we are able to estimate the imputation error. Evaluation is performed on two molecular descriptor datasets generated from a diverse selection of pharmaceutical fields with artificially introduced missing values ranging from 10% to 30%. The experimental result demonstrates that missing values has a great impact on the effectiveness of imputation techniques and our method MiFoImpute is more robust to missing value than the other ten imputation methods used as benchmark. Additionally, MiFoImpute exhibits attractive computational efficiency and can cope with high-dimensional data.*

## 1. INTRODUCTION

The nature of molecular descriptor data complicates the development of highly accurate predictive models. Molecular descriptor data are typically inconsistently gathered. These empty or unanswered values in data sets are named missing values (data), and are of a problem most researchers face. Missing data may occur from various reasons. For instance, accidentally or some molecules descriptor generator are fail to produce descriptor data. Imputation of missing values (MV) is often a necessary step in data analysis. MV imputation remains a necessary key step in data preprocessing. Since many down-stream analyses require a complete data set for implementation, MV imputation is a common practice.

Many established procedures of analysis require fully observed molecular descriptor datasets without any missing values. However, this is infrequently the case in pharmaceutical and biological research today. The continuous development of new and improved measurement techniques in these fields provides data analysts with challenges prompted not only by high-dimensional multivariate descriptor data where the number of descriptors may greatly exceed the number of observations where continuous descriptors are present.

Many MV imputation methods have been developed in the literature. MV imputation methods generally belong to two categories. In the first category, expression information of a missing entry is borrowed from neighbouring descriptors whose closeness is determined by a distance measure (e.g., correlation, Euclidean distance). Those methods are k nearest neighbours [1], generalized boosted model [2], Locally Weighted Linear Imputation [3], Mean Imputation [4], SVD Imputation [5], SVT Imputation [6], Approximate SVT Imputation [7].All these methods are based on the fact that molecular descriptor do not function individually, but are usually highly correlated with co-regulated descriptor. For the second category, dimension reduction techniques are applied to decompose the data matrix and iteratively reconstruct the missing entries. Those methods are Bayesian Principal Component Analysis(BPCA)[8], Probabilistic PCA(PPCA)[9] and LLSimpute[10],. The above imputation methods are restricted to one type of variable. Furthermore, all these methods make assumptions about the distribution of the data or subsets of the variables, leading to questionable situations, e.g. assuming normal distributions.

Our motivation is to introduce a method of imputation which can handle any type molecular descriptor data and makes as few as possible assumptions about structural aspects of the data. Random forest [11] is able to deal with real valued-type data and as a non-parametric method it allows for interactive and non-linear (regression) effects. We take up the missing data problem using an iterative imputation scheme by training an RF on observed values in a first step, followed by predicting the missing values and then go on iteratively. We choose RF because it is known to perform very well under barren conditions like high dimensions, complex interactions and non-linear data structures. Due to its accuracy and robustness, RF is well suited for the use in applied research often recalling such conditions.

Here we compare our method with two categories of imputation methods are mentioned above paragraph. These imputation methods applied on molecular descriptor datasets. Comparisons are performed on two molecular descriptor datasets generated from Padel-Descriptor generator [12] and using different proportions of missing values. Missing values are indicated by NAs in R [18] We show that our approach is competitive to or outperforms the compared methods on the used datasets irrespectively of the variable type composition, the data dimensionality, the source of the data or the amount of missing values.

In some cases, the decrease of imputation error is up to 50%. This performance is typically reached within only a few iterations which makes our method also computationally attractive. The NRMSE and NMAE error estimates give a very good approximation of the true imputation error having on average a proportional deviation of no more than 10–15%. In addition, our method needs no tuning parameter, and hence is easy to use.

## 2. MATERIALS AND METHODS

### 2.1 The Data Sets

We investigate the following two molecular descriptor datasets. The first ,the molecular descriptors of Oxazolines and Oxazoles derivatives [15-16] based H37Rv inhibitors. The dataset covers a diverse set of molecular descriptors with a wide range of inhibitory activities against H37Rv. This molecular Descriptor data set includes 100 observations with 254 descriptors. The second ,the molecular descriptors of Thiolactomycin and Related Analogues [13] based H37Rv inhibitors. The dataset covers a diverse set of molecular descriptors with a wide range of inhibitory activities against H37Rv. This molecular Descriptor data set includes 200 observations with 255 descriptors.

### 2.2 Algorithmic Approach

Let's assume $X = \{X_1, X_2, \ldots, X_p\}$ to be a $n \times p$-dimensinal descriptor dataset matrix. We propose method using an Random Forest (RF) to impute the missing values due to its prior mentioned advantages as a regression method. The RF algorithm has a predefined function to handle missing values by weighting the number of the observed values in a variable with the RF togetherness after being trained on the initially mean imputed descriptor dataset [14]. After all, this method requires a complete dependent variable for training the forest.

In place of, straightforwardly predict the missing values using an RF trained on the observed parts of the descriptor dataset. For an arbitrary descriptor $X_t$ containing missing values at entries $j_{mis}^{(t)} \subseteq \{1, \ldots, n\}$ we can come apart the dataset into four parts:

(a) The observed values of descriptor $X_t$, indicated by $y_{obs}^{(t)}$;
(b) The missing values of descriptor $X_t$, indicated by $y_{mis}^{(t)}$;
(c) The descriptors other than $X_t$ with observations $j_{obs}^{(t)} = \{1, \ldots, n\} \setminus j_{mis}^{(t)}$ indicated by $X_{obs}^{(t)}$; and
(d) The descriptors other than $X_t$ with observations $j_{mis}^{(t)}$ indicated by $X_{mis}^{(t)}$.

Indicate that $X_{obs}^{(t)}$ is commonly not completely observed since the index $j_{obs}^{(t)}$ corresponds to the observed values of the descriptor $X_t$. $X_{mis}^{(t)}$ is commonly not completely missing.

To start , build an initial guess for the missing values in $X$ using mean imputation or another imputation method. Then, sort the descriptors $X_t$, $t = 1, \ldots, p$ according to the amount of missing values beginning with the lowest amount. For every descriptor $X_t$, the missing values are imputed by first fitting an RF with response $y_{obs}^{(t)}$ and predictors $X_{obs}^{(t)}$ then, predicting the missing values $y_{mis}^{(t)}$ by applying the trained RF to $X_{mis}^{(t)}$. The imputation method is repeated until a termination criterion is met. Algorithmic approach of missing forest Imputation (MiFoImpute) method is given below.

*Algorithm:* Impute missing values with RF

*INPUT:* $X$ an $n \times p$ descriptor dataset matrix, termination criterion $\gamma$
*i)* Build initial guess for missing values;
*ii)* $l \leftarrow$ vector of sorted indices of columns in $X$ with respect to increasing amount of missing values;

*iii)* **while** not $\gamma$ **do**

*iv)* $X_{old}^{(imp)} \leftarrow$ store previously imputed matrix;

*v)* **for** $t$ **in** $k$ **do**

*vi)* Fit a random forest: $y_{obs}^{(t)} \sim X_{obs}^{(t)}$;

*vii)* Predict $y_{mis}^{(t)}$ using $X_{mis}^{(t)}$

*viii)* $X_{new}^{imp} \leftarrow$ update imputed matrix, using predicted $y_{mis}^{(t)}$;

*ix)* **end for**

*x)* update $\gamma$.

*xi)* **end while**

*xii)* **return** the imputed matrix $X^{imp}$

The termination criterion $\gamma$ is met as soon as the difference between the newly imputed data matrix and the previous one increases for the first time with respect to both variable types, if present. Here, the difference for the set of descriptor variables $N$ defined as

$$\Delta_N = \frac{\sum_{i \in N}(X_{new}^{imp} - X_{old}^{imp})^2}{\sum_{i \in N}(X_{new}^{imp})^2},$$

## 2.3 Performance Measure

After imputing the missing values, the performance is evaluated using the normalized root mean squared error (NRMSE) [14] for the descriptor variables which is defined by

$$\text{NRMSE} = \sqrt{\frac{\text{mean}((X^{true} - X^{imp})^2)}{\text{var}(X^{true})}},$$

Where $X^{true}$ is the complete descriptor data matrix and $X^{imp}$ the imputed descriptor data matrix. use mean and var as short notation for empirical mean and variance computed over the missing values only. Good performance leads to a value close to 0 and bad performance to a value around

1. To evaluate the precision of imputation, the normalized mean absolute error (NMAE) [17] is used and its value at variable $X_k$ is calculated as follows:

$$NMAE_k = \frac{1}{n_k^{mis}} \sum_{i=1}^{n_k^{mis}} \frac{(X_k^{imp} - X_k^{true})}{X_k^{max} - X_k^{min}},$$

Where $n_k^{mis}$ is the number of missing values at $X_k$, $X_k^{true}$ and $X_k^{imp}$ denote the true value and imputed value of the missing data respectively, $X_k^{max}$ and $X_k^{min}$ are the maximum and minimum value at $X_k$. The value of NMAE on the whole datasets takes the average over all the descriptor variables.

## 3. EXPERIMENTAL RESULTS AND ANALYSIS

### 3.1 Generation of missings in the dataset

Given a two molecular descriptor datasets, a pattern of missing entries (NA) is produced randomly on a matrix of the size of the data set with a pre-specified proportion of the missings. The proportion of missing entries may vary. We use the random uniform distribution for generating missing positions and the proportion's range at 10%, 20% and 30% of the total number of entries.

### 3.2 Evaluation of results

Since the missings are generated separately, we can evaluate the quality of imputation by comparing the imputed values with those generated at the stage of missing entries (NA) is produced. We use the NRMSE and NMAE, to measure the performance of an algorithm. The imputation results for MiFoImpute on the Oxazolines and Oxazoles derivatives and Thiolactomycin derivatives molecular descriptor data have to be treated. Figure 1 and Figure 2 presents a comparative study between the performances (NRMSE) of various imputation methods over two descriptor datasets. Each curve in the figure represents the results from one imputation method.

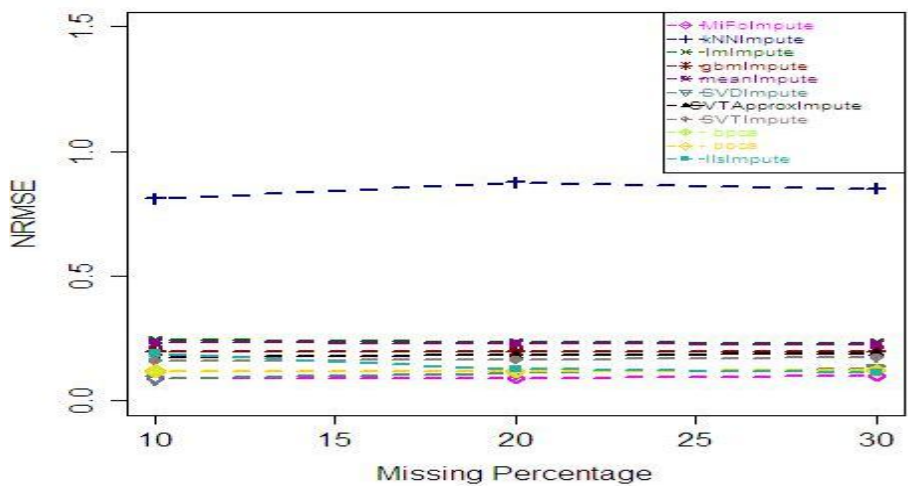

Figure. 1. Performance of the eleven imputation methods on Oxazolines and Oxazoles derivatives descriptor data. The percentage of entries missing in the complete dataset and the NRMSE of each missing value estimation method are shown in the horizontal and vertical axes, respectively Figure. 1 shows among all other imputation methods, the MiFoImpute method gives comparable NRMSE values. From this Figure. 1, we see that when the percentage of missing values in the data set is 20%, the MiFoImpute achieves best results. When the percentage of the missing values reaches 30%, the NRMSE of the MiFoImpute is little and also achieves best results than bpca ,SVDImpute and ppca impute methods. This shows that MiFoImpute method is comparable with if not better than the previous methods on this data set. Performance's of KNNImpute method is worst of all the eleven imputation methods.

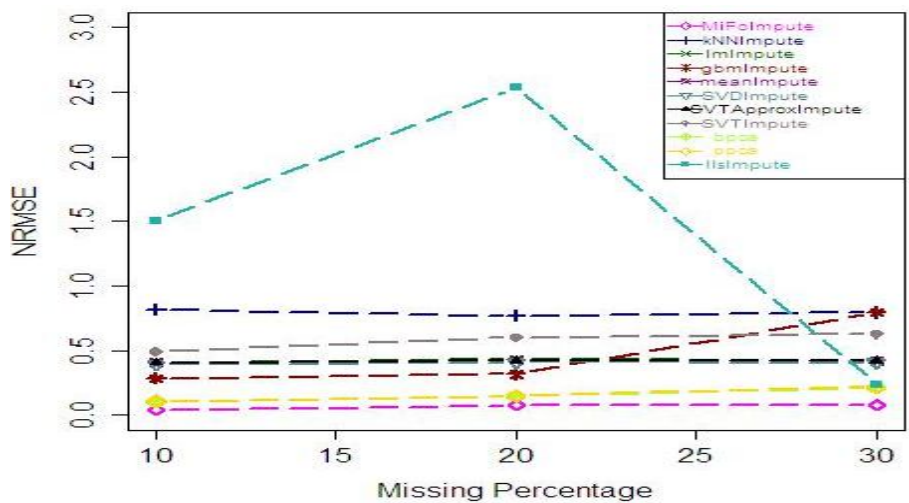

Figure. 2. Performance of the eleven imputation methods on Thiolactomycin derivatives descriptor data. The percentage of entries missing in the complete dataset and the NRMSE of each missing value estimation method are shown in the horizontal and vertical axes, respectively From Figure. 2, we see that MiFoImpute method starts to outperform the other impute methods when the missing rate is increased especially on the Thiolactomycin descriptor data set. Generally, the MiFoImpute performs stable across the missing data. For example, all the other methods give an estimate performance with NRMSE between 2.538303 and 0.1442081 for 15% missing, whereas MiFoImpute gives 0.07245663. Consequently, MiFoImpute impute method performs robustly as the percentage of the missing values increase. MiFoImpute achieves best results than bpca and ppca. Performance's of llsImpute and KNNImpute methods are worst of all the eleven imputation methods.Figure. 3 and Figure. 4 demonstrates normalized mean absolute error (NMAE) measure to compare various imputation methods over two descriptor datasets. Each curve in the figure represents the results from one imputation method.

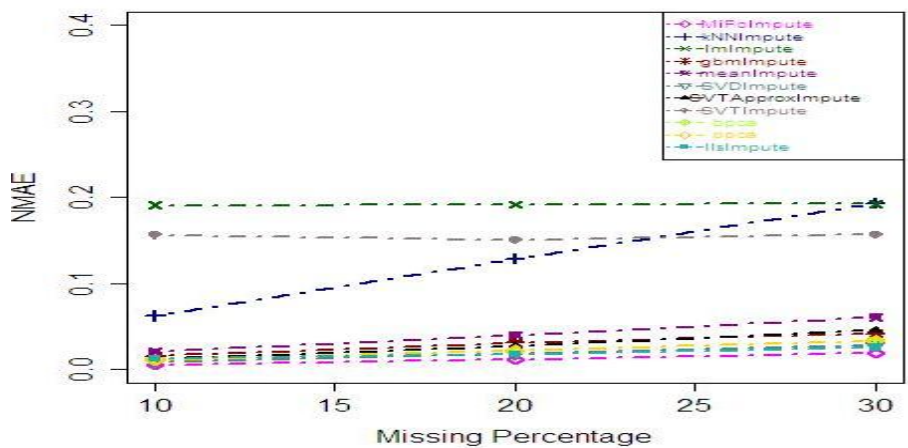

Figure. 3. Performance of the eleven imputation methods on Oxazolines and Oxazoles derivatives descriptor data. The percentage of entries missing in the complete dataset and the NMAE of each missing value estimation method are shown in the horizontal and vertical axes, respectively

First, different noise levels have different impacts on imputation accuracy. Generally speaking, the NMAE increases with the level of missing values for all the methods. This is understandable because with more missing values introduced into the datasets, more negative effects will be brought to the imputation results. Nevertheless, although the noise will deteriorate the imputation accuracy, when comparing the results from three missing percentage. However, when level missing percentage is relatively high, the introducing of more missing values will deteriorate the imputation results dramatically. Take Figure 3 in which the missing percentage equals to 30% as an example, when missing percentage increases from 10% to 20%,the error of MiFoImpute increases slightly from 0.00551459 to 0.01150088. But when missing percentage reaches to 30%, the error degrades to 0.01913564. For the Oxazolines and Oxazoles derivatives descriptor data , when comparing different methods, MiFoImpute achieve better accuracy than the other three methods and their accuracy difference is indiscernible. bpca and ppca appear to be the second best methods. lmImpute and knnImpute methods are worst of all the eleven imputation methods.

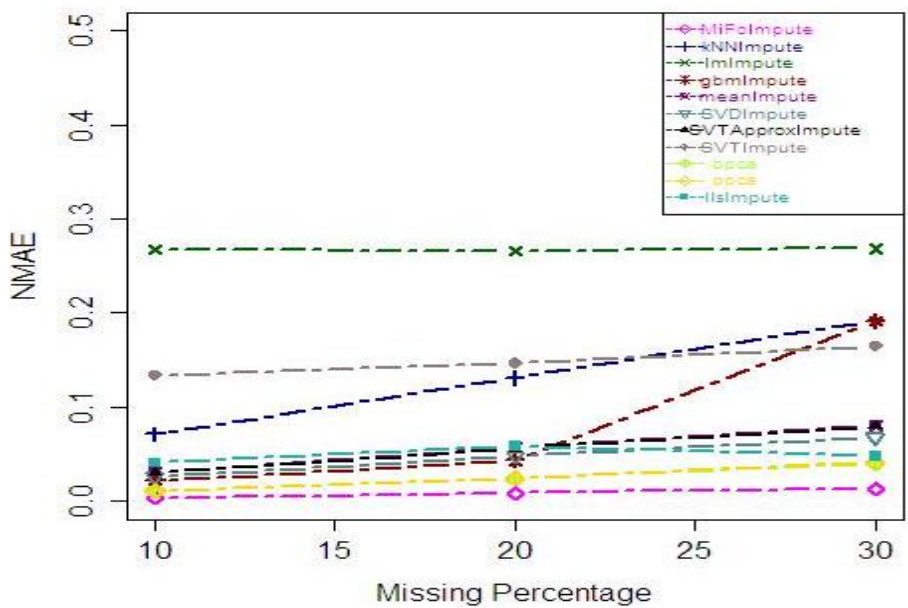

Figure. 4. Performance of the eleven imputation methods on Thiolactomycin derivatives descriptor data. The percentage of entries missing in the complete dataset and the NMAE of each missing value estimation method are shown in the horizontal and vertical axes, respectively Take Figure 4 in which missing percentage increases from 10% to 20%, the error(NMAE) of MiFoImpute increases slightly from 0.003833351 to 0.008735986. But when missing percentage reaches to 30%, the error degrades to 0.01336493. For the Thiolactomycin derivatives descriptor data , when comparing different methods, MiFoImpute achieve better accuracy than the other three methods and their accuracy difference is indiscernible. bpca and ppca appear to be the second best methods. lmImpute, gbmImpute and knnImpute methods are worst of all the eleven imputation methods.

### 3.3 Computational Efficiency

To determine the computational cost of MiFoImpute by comparing the runtimes of imputation on the previous two descriptor datasets.

Table 1 shows the runtimes in seconds of all methods on the analysed descriptor datasets.

| Dataset | n | p | MiFoImpute | knnImpute | lmImpute | gbmImpute | meanImpute | SVDImpute | SVTApproxImpute | SVTImpute | bpca | ppca | llsImpute |
|---|---|---|---|---|---|---|---|---|---|---|---|---|---|
| Oxazoline and Oxazoles | 100 | 254 | 145.53 | 1.317 | 0.773 | 213.67 | 0.567 | 0.287 | 0.047 | 0.31 | 3.477 | 0.463 | 329.13 |
| Thiolactomycin | 200 | 255 | 264.15 | 2.373 | 0.76 | 629.63 | 0.033 | 0.887 | 0.123 | 1.047 | 4.03 | 2.43 | 1725.5 |

We can see SVTApproxImpute that is by far the fastest method for the first dataset. meanImpute is the fastest method for the second dataset. However, MiFoImpute runs considerably faster than gbmImpute and the llsImpute for both datasets. There are two possible ways to speed up computation. The first one is to reduce the number of trees grown in each forest. In all comparative studies, the number of trees was set to 100 which offers high precision but increased runtime.

Table 2. Average imputation error (NRMSE/NMAE in percent) and runtime (in seconds) with different numbers of trees $(n_{tree})$ grown in each forest and descriptors tried $(m_{try})$ at each node of the trees. Here, we consider the Oxazolines and Oxazoles derivatives descriptor dataset with artificially introduced 10% of missing values.

| $m_{try}$ | $n_{tree}$ | | | | |
|---|---|---|---|---|---|
| | 10 | 50 | 100 | 250 | 500 |
| 1 | 11.78/1.01 6.28s | 10.82/0.91 20.35s | 10.66/0.90 29.47s | 10.63/0.89 73.06s | 10.58/0.89 206.35s |
| 2 | 10.21/0.89 5.43s | 9.91/0.81 24.11s | 9.77/0.80 36.26s | 9.64/0.79 95.42s | 9.65/0.79 315.96s |
| 4 | 9.40/0.76 6.41s | 9.06/0.71 36.34s | 8.82/0.69 46.66s | 8.76/0.69 125.71s | 8.78/0.68 262.99s |
| 8 | 8.61/0.67 6.41s | 8.06/0.60 52.10s | 7.93/0.60 67.00s | 7.91/0.59 185.39s | 7.90/0.59 492.90s |
| 16 | 7.81/0.59 12.36s | 7.29/0.52 82.82s | 7.25/0.52 148.07s | 7.28/0.51 378.73s | 7.19/0.51 799.75s |

In Table 2 and Table3, we can see that changing the number of trees in the forest has a stagnating influence on imputation error, but a strong influence on computation time which is approximately linear in the number of trees. The second one is to reduce the number of descriptors randomly selected at each node $(m_{try})$ to set up the split

International Journal on Computational Sciences & Applications (IJCSA) Vol.3, No4, August 2013

Table 3. Average imputation error (NRMSE/NMAE in percent) and runtime (in seconds) with different numbers of trees ($n_{tree}$) grown in each forest and descriptors tried ($m_{try}$) at each node of the trees. Here, we consider the Thiolactomycin derivatives descriptor dataset with artificially introduced 10% of missing values.

| $m_{try}$ | $n_{tree}$ | | | | |
|---|---|---|---|---|---|
| | 10 | 50 | 100 | 250 | 500 |
| 1 | 22.15/1.27 7.02s | 18.89/1.15 21.59s | 19.04/1.17 43.72s | 18.67/1.16 117.17s | 17.25/1.11 361.71s |
| 2 | 14.85/0.88 12.63s | 10.62/0.72 31.93s | 10.14/0.71 58.39s | 10.27/0.70 233.92s | 10.20/0.71 480.80s |
| 4 | 9.96/0.69 17.22s | 7.11/0.55 55.03s | 6.97/0.54 117.03s | 6.90/0.54 409.13s | 7.29/0.54 1153.99s |
| 8 | 6.39/0.50 13.68s | 5.34/0.45 65.15s | 5.58/0.44 116.94s | 5.21/0.43 435.64s | 5.21/0.43 965.33s |
| 16 | 6.53/0.44 16.75s | 4.15/0.37 94.58s | 4.53/0.38 151.22s | 4.35/0.37 831.35s | 4.33/0.371 134.75s |

Table 2 and Table 3 shows that increasing $m_{try}$ has limited effect on imputation error, but computation time is strongly increased. Note that for $m_{try} = 1$ we no longer have a Random Forest, since there is no more choice between descriptors to split on. This leads to a much higher imputation error, especially for the cases with low numbers of bootstrapped trees. We use for all experiments $\sqrt{p}$ as a default value.

## 4. CONCLUSIONS

We presented the MiFoImpute algorithm, as an alternate and practical approach to the data imputation problem , it can handle multivariate data consisting of molecular descriptors MiFoImpute has no need for tuning parameters nor does it require assumptions about distributional aspects of the data. Different missing values imputation methods were examined. Each method was tested under two molecular descriptor data sets . By observing the behaviour of the different imputation methods at different missing levels, we drew the conclusion that missing values has great negative effects on imputation methods, especially when the missing level is high. NRMSE and NMAE measures were used to measure the performance of the algorithms. Comparative studies have shown that MiFoImpute performs quite well in comparison with other ten popular imputation methods in the presence of missing values The goal of MiFoImpute method is to provide an accurate way of estimating missing values in order to minimally bias the performance of imputation methods. MiFoImpute can be applied to high-dimensional datasets where the number of variables may greatly exceed the number of observations to a large extent and still provides excellent imputation results.

## ACKNOLDGEMENTS

We gratefully thank to the Department of Computer Science Mangalore University, Mangalore India for technical support of this research.

**Authors**

Doreswamy received B.Sc degree in Computer Science and M.Sc Degree in Computer Science from University of Mysore in 1993 and 1995 respectively. Ph.D degree in Computer Science from Mangalore University in the year 2007. After completion of his Post-Graduation Degree, he subsequently joined and served as Lecturer in Computer Science at St. Joseph's College, Bangalore from 1996-1999.Then he has elevated to the position Reader in Computer Science at Mangalore University in year 2003. He was the Chairman of the Department of Post-Graduate Studies and research in computer 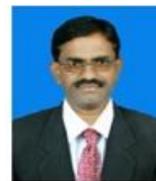 science from 2003-2005 and from 2009-2008 and served at varies capacities in Mangalore University at present he is the Chairman of Board of Studies and Associate Professor in Computer Science of Mangalore University. His areas of Research interests include Data Mining and Knowledge Discovery, Artificial Intelligence and Expert Systems, Bioinformatics ,Molecular modelling and simulation Computational Intelligence,Nanotechnology, Image Processing and Pattern recognition. He has been granted a Major Research project entitled "Scientific Knowledge Discovery Systems (SKDS) for Advanced Engineering Materials Design Applications" from the funding agency University Grant Commission, New Delhi , India. He has been published about 30 contributed peer reviewed Papers at national/International Journal and Conferences. He received SHIKSHA RATTAN PURASKAR for his outstanding achievements in the year 2009 and RASTRIYA VIDYA SARASWATHI AWARD for outstanding achievement in chosen field of activity in the year 2010.

Chanabasayya.M. Vastrad received B.E. degree and M.Tech. degree in the year 2001 and 2006 respectively. Currently working towards his Ph.D Degree in Computer Science and Technology under the guidance of Dr. Doreswamy in the Department of Post-Graduate Studies and Research in Computer Science, Mangalore University. 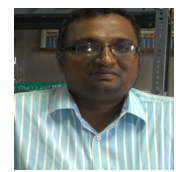